\documentclass{aa}
\usepackage{graphics}
\usepackage{txfonts}
\begin{document}
 
\title{Menzel 3: dissecting the ant} 
\author{\it Miguel Santander-Garc\'\i a\inst{1}, Romano L.\ M. Corradi\inst{2}, Bruce Balick\inst{3},
\and Antonio Mampaso\inst{1}} 

\institute{
	 Instituto de Astrof\'\i sica de Canarias, 38200 La Laguna, 
        Tenerife, Spain
         \\e-mail: miguelsg@iac.es; amr@iac.es
        \and
	 Isaac Newton Group of Telescopes, Ap.\ de Correos 321,
	 38700 Sta. Cruz de la Palma, Spain
         \\e-mail: rcorradi@ing.iac.es
	\and 
         Department of Astronomy, University of Washington, Seattle, Washington 98195-1580, USA
         \\e-mail: balick@astro.washington.edu
	} 
\offprints{M. Santander-Garc\'\i a}
%\date{Received ...; accepted ...}     
%\date{\today}
 
\abstract{
The structure and kinematics of the bipolar nebula Mz~3 have been investigated by means of HST,
CTIO and ESO images and spectra. At least four distinct outflows have been identified which, from
the inside to the outside, are the following: a pair of bright bipolar lobes, two opposite highly
collimated column-shaped outflows, a conical system of radial structure, and a very dim, previously
unnoticed, low-latitude and flattened (ring-like) radial outflow.

A simple Hubble-law describes the velocity field of the ballisticaly expanding lobes, columns and rays,
suggesting that their shaping has being done at very early stages of evolution, in a sort of eruptive 
events  with increasing degree of collimation and expansion ages ranging from $\sim$600 for the
inner structures to $\sim$1600 years (per kpc to the nebula) for the largest ones. 

\keywords{planetary nebulae: Mz~3 -- interstellar medium: kinematics and dynamics} 
}

\authorrunning{Santander-Garc\'\i a et al.}
\titlerunning{Menzel 3: dissecting the ant}
\maketitle   
 
\section{Introduction} 
Menzel 3 (Mz~3, also named as PN G331.7-01.0 or He2-154) is one of the most strikingly beautiful
and complex bipolar nebula, named the Ant for its characteristic morphology (Menzel \cite{Me22}). The
nebula shows several components with different degrees of collimation, which together with its
unusual spectrum made it meriting the nickname of ``The Chamber of Horrors" of planetary nebulae
(Evans \cite{Ev59}).

Currently, the nature of the central star of Mz~3 is under discussion. While it has been classicaly
classified as a young planetary nebula (PN), the near-IR (J-K)$_0$ vs.\ (I-J)$_0$ diagram by Schmeja \&
Kimeswenger (\cite{SK01}) shows that it lies far away from the classical PN region and very close to the locus
occupied by symbiotic Miras. This and a complex stellar spectrum with many iron emission lines suggest that
Mz~3 might have a symbiotic binary central star, surrounded
by a high density equatorial gas disc (Smith \cite{Sm03}). 
Part of the complexity of Mz~3 resides in its distinct multiple outflows. Only few bipolar PNe and nebulae
around symbiotic Miras show multiple bipolar lobes and jets, a phenomenon which looks instead rather common
for pre-PNe (Sahai \& S\'anchez-Contreras \cite{SS02}). Examples are M2-9 (also suspected to have a symbiotic
nucleus), He2-104 (a genuine symbiotic Mira, Corradi et al. \cite{Co01}), as well as the following objects
classified as PNe: He2-320, He2-86, He3-401, He2-437, M2-46 and M3-28 (Manchado et al. \cite{Ma96}). 

In this paper, we analyse images and high resolution spectra of Mz~3 by means of a spatiokinematical modeling
which allows us to determine the geometrical structure, velocity field and orientation in the sky 
of the different outflows. The motivation for this is clear: the scientific method is built on identifying
patterns, unusual as they may be, before trying to explain them.

Preliminary spatiokinematical modeling of Mz~3 were presented by Santander-Garc\'\i a (\cite{Sa04}).

\begin{figure*} 
\center
\resizebox{3.78cm}{!}{\includegraphics{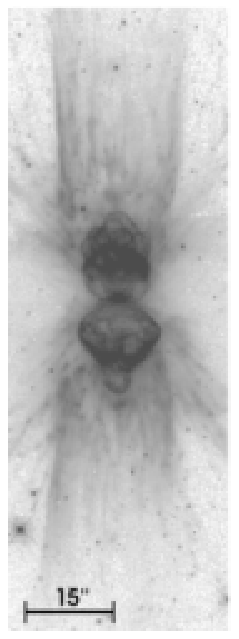}} 
\resizebox{5.7cm}{!}{\includegraphics{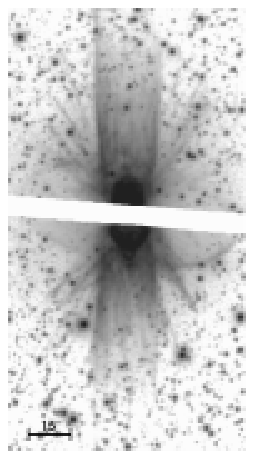}} 
\resizebox{5.7cm}{!}{\includegraphics{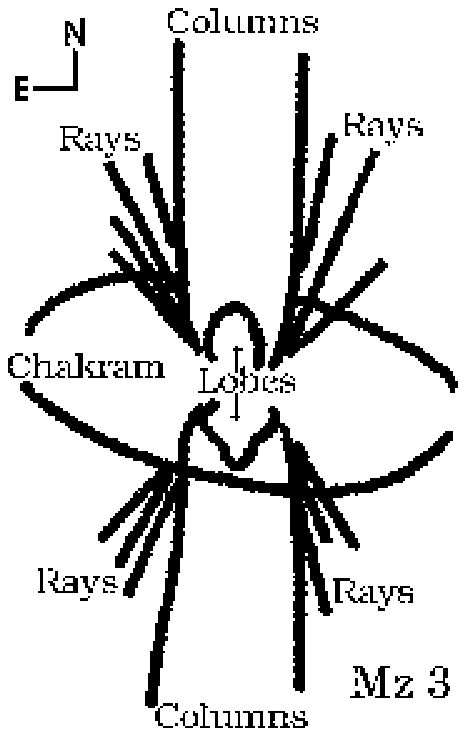}} 
\caption{{\bf Left} \& {\bf Middle}: HST \& ESO H$\alpha$+[N{\sc ii}]\   images. The white bar in the ESO image corresponds
to the gap between the 2 CCD of the mosaic camera SUSI2. {\bf Right}: Sketch of Mz~3,
showing the various nebular components discussed in the text.}
\label{F1}
\end{figure*}

\begin{figure*}
\center
\resizebox{6.0cm}{!}{\includegraphics{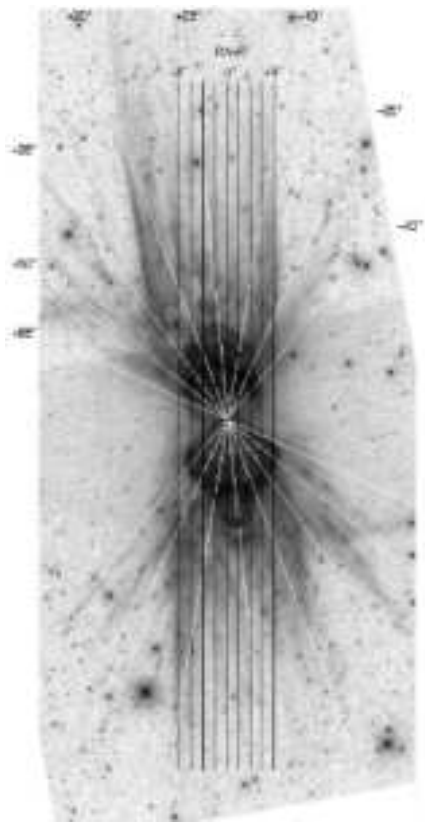}}  
\caption{Echelle long-slit location display over Mz~3. CTIO slit positions are indicated by black
lines, whereas white ones indicate ESO slit positions.} 
\label{F0}
\end{figure*}

\begin{figure*}[!]
\center
\resizebox{2.64cm}{!}{\includegraphics{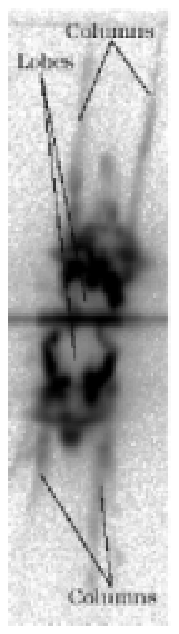}}
\resizebox{4.125cm}{!}{\includegraphics{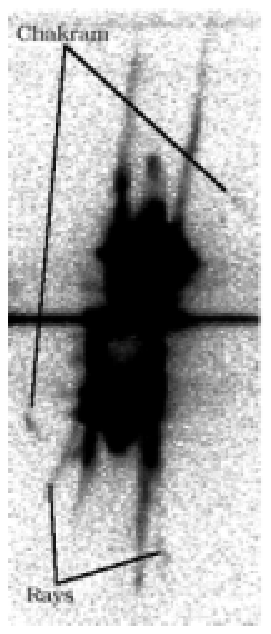}}
\caption{The [N{\sc ii}]\  ESO spectrum at P.A.=5$^\circ$, shown with two different contrasts. The association of
velocity features with the different nebular components is indicated. Each frame is
100$''$ along $y$ axis. Left and right frame are 275 and 430 km s$^{-1}$, respectively, along $x$
axis (both frames have the same scale). Northern side is at the top.} 
\label{F2}
\end{figure*}

\section{Observations and Data reduction}

Images of Mz~3 were obtained using the Wide Field Planetary Camera 2 (``WFPC2") on the Hubble Space Telescope
(``HST") with 0$''$.1 pixel$^{-1}$ in 1997 (GO6502) and 1998 (GO6856), and with 0$''$.045 pixel$^{-1}$ in 2001
(GO9050). For this paper we use only the images obtained through the F656N filter (center
wavelength/bandpass = 6564/22 \AA) and the F658N filter (6590/29 \AA) from the HST archive. Pointing
positions and exposure times can be obtained from the HST archive. The F658N filter transmits the [N{\sc ii}]\ 
line with 78\%  efficiency and the H$\alpha$\  line with about 4\%  efficiency, depending on the Doppler shift
of the target and its internal kinematics. Generally this is not a problem since the H$\alpha$\  and
[N{\sc ii}]\  images are very similar in appearance and intensity except within about 1'' of the nucleus where
the H$\alpha$\  image shows a stronger central peak than does [N{\sc ii}]\  (the flux of the central star is over
50 times greater in H$\alpha$\  than it is in [N{\sc ii}]\ ). The image in filter F658N is presented in Fig.~\ref{F1}.

A slightly deeper image (300 s) was obtained on February 1, 2001, at the 3.5 m New Technology
Telescope (NTT) at ESO with its imager SUSI2. H$\alpha$\  and [N{\sc ii}]\  $\lambda\lambda$6548,6583 were
isolated with the aid of the Hal\#884 filter, with central wavelength/FWHM 6555.3/69.8 \AA. The
seeing was 0$''$.7. The spatial scale on the detectors, 2 EEV44-80 CCD (2048$\times$4096 pixels$^2$
of 15$\mu$m size each), is 0$''$.24 pixel$^{-1}$ after a 3$\times$3 pixels binning. This image is also presented in
Fig.~\ref{F1}.
 
Long-slit echelle spectra were obtained at the 4-m Blanco Telescope at CTIO, on July 11, 1998. Nine
exposures, each of 300~s, at P.A.=$0^\circ$ and offsets from -8$''$ (E) to +8$''$ (W) from the
central star with steps of 2$''$, were obtained. They cover a narrow spectral region including the
H$\alpha$\  and [N{\sc ii}]\  $\lambda\lambda$6548,6583 nebular line emission. The spatial sampling is 0$''$.27
pixel$^{-1}$, while the velocity resolution is 3.75 km s$^{-1}$. Slit width was 1$''$.

A second
set of high-resolution spectra was secured with the EMMI instrument and its Echelle grating at the
NTT on May 12, 2000. It consists of 8 exposures, each of 1000~s, spanning from P.A.=$-40^\circ$ to
P.A.=$+65^\circ$ with $15^\circ$ increments and centered on the central star. The Ha\#596 filter was
used to isolate the order containing H$\alpha$\  and the nitrogen doublet. The spatial sampling and
velocity dispersion are 0$''$.27 pixel$^{-1}$ and 1.88 km s$^{-1}$, respectively. The slit width was
1$''$.2, and seeing between 1$''$.3 and 1$''$.7. All the slit positions of the Blanco Telescope and
NTT spectra are shown in Fig.~\ref{F0}.

Additionally, a long-slit spectrum of Mz~3 was obtained with STIS at the HST on June 23, 2002.
Exposure time was 217~s, and the slit was oriented at P.A.=$0^\circ$ through the central star. The
spatial scale is 0''.05 pixel$^{-1}$, and the velocity dispersion is 23.44 km~s$^{-1}$. This
spectrum also cover H$\alpha$\  and [N{\sc ii}]\  $\lambda\lambda$6548,6583.

Data were reduced using standard IRAF packages.

\section{Morphology and spectra}

The images of Mz~3 are presented in Fig.~\ref{F1}, while the long-axis, central spectra are shown in
Fig.~\ref{F2}. The nebula exhibits a complex morphology with the evidence for several components
(Meaburn \& Walsh 1985) that we attempt to study separately. In particular, we can distinguish at
least four morphological components as indicated in the sketch of Fig.~\ref{F1}. 

The brightest features are a pair of bipolar {\it lobes} emerging from the central star, and extending out
to
about 15$''$ in opposite directions. They look, both in the images and in the
spectra, like bubbles with protrusions at the top and bottom. In the HST image, there is some
evidence for a second pair of smaller, inner lobes (see Fig.~\ref{F3}), but they are not easily distinguished from the
larger lobes in the images and in the kinematical data, so that we did not try to model them. At the
base of the lobes, the surface brightness is
clearly enhanced in all emission lines, suggesting that very large gas densities are present. The region is briefly
discussed in section 4.2. 

The second brightest feature is the two symmetrical, long tubular protuberances emerging from the
central region, that we call {\it columns}. They are easily recognized in the spectra as pairs of 
inclined parallel lines.
 
The {\it rays} are fainter radial features extending from the central region in a sector within
$\sim$45$^\circ$ from the symmetry axis of the lobes and columns. In the spectra, they appear as dim
lines, to the left and right of the columns, emanating from the central star. 

An even more puzzling feature is what we have called the {\it chakram} because of some similarity
(after the modeling described below) with an ancient disk-like Hindu throwing weapon. In the ESO
image, it appears as a very faint filled ellipse which major axis is tilted by $\sim$80$^\circ$ with
respect to the long axis of the other outflows. This structure has been unnoticed in previous
studies of the nebula. It corresponds to very faint, arc-shaped, and high-velocity features near the 
upper-right and lower-left corners of all spectra (see Fig.~\ref{F2}), showing decreasing radial
velocities at larger distances from the star. Redman et al. (\cite{Re00}) noticed these high velocity
features in their spectra, but attributed them to the polar protrusions of the lobes, as they lacked
of spectra extending to regions outside the inner bipolar lobes and columns. 
 
\section{Spatiokinematical modeling} 

The wavelengths of the long-slit echelle spectra were transformed into
radial velocities for the [N{\sc ii}]\ $\lambda$ 6583 line. We have
chosen this line instead of H$\alpha$\ because its distribution is
similar to, its surface brightness is slightly stronger than, and its
thermal broadening is much smaller than that of H$\alpha$\ over the
nebula (apart from the core).

In order to model the images and spectra, we used a spatiokinematical
model based on the assumptions of (a) axial-symmetry,  (b) that
the streamlines of gas are radial, and (c) each particle is expanding
with a velocity proportional to the present distance from the star
($v_{exp} \propto r$). This second assumption corresponds to a
``Hubble law'', and is found to provide a fair representation of a
large fraction of bipolar nebulae (see Corradi \cite{Co04}).

The code, written in IDL, produces an axisymmetric, two-dimensional
model which is then scaled according to the distance from the Earth
and rotated into three dimensions about its symmetry axis and inclined
to the plane of the sky to allow direct comparison with the
images. Spatial velocities at each point of the nebula are also
computed assuming the previously mentioned Hubble law (except for the
chakram, see below), and then transformed to radial velocities
according to the adopted orientation parameters. Then synthetic
position-velocity plots for several slit positions and offsets are
generated. These are compared with our echelle spectra.

The model fit to the data is performed visually after allowing the
different geometrical and kinematical parameters to vary over a large
range of values. It should also be stressed that the various
parameters are not entirely independent of each other, so that varying
one parameter value within its acceptable range implies that the other
parameters must be adjusted accordingly.
 
\subsection{Lobes and nucleus}

 \begin{figure*}[!]
\center
\resizebox{3.0cm}{!}{\includegraphics{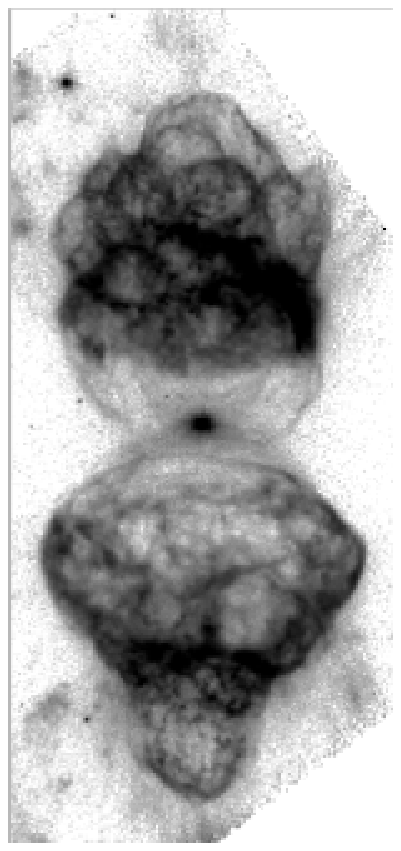}}
\resizebox{3.0cm}{!}{\includegraphics{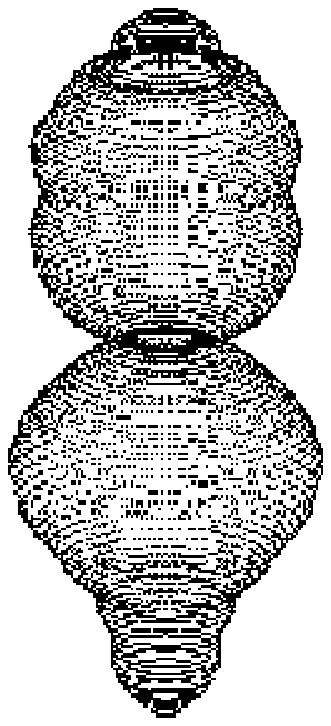}}
\caption{{\bf Left}: The H$\alpha$\ +[N{\sc ii}]\   HST image of the lobes of Mz~3. {\bf Right}: Adopted
model.} 
\label{F3}
\end{figure*}

\begin{figure*}[!]
\center
\resizebox{9.0cm}{!}{\includegraphics{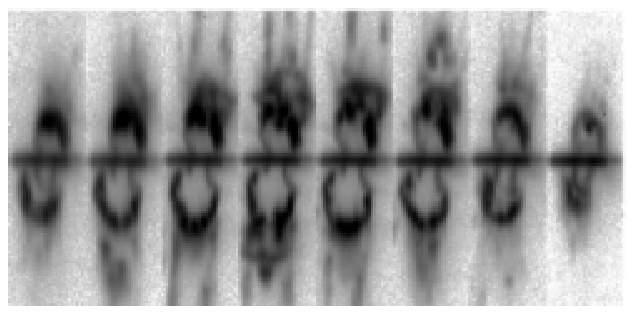}}
\resizebox{9.0cm}{!}{\includegraphics{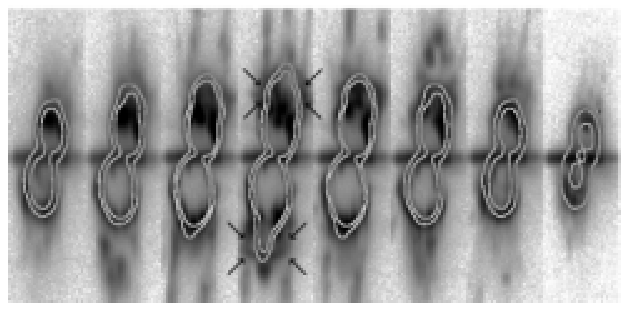}}
\caption{{\bf Top}: The ESO [N{\sc ii}]\  echelle spectra. Frames correspond, from left to right, to P.A. from $-40^\circ$
to $+65^\circ$ in increments of $15^\circ$ and no offset. {\bf Bottom}: The same spectra with the spatio-kinematical model
of the lobes superimposed. Model lines have been artificially broadened to facilitate the
display and to improve the S/N ratio. The main velocity deviations from this
Hubble-like flow model are indicated by
arrows. {\bf Note}: Each frame is 50$''$ tall and 200
km s$^{-1}$ wide. Northern side is up.} 
\label{F4}
\end{figure*}

\begin{figure*}[!]
\center
\resizebox{6.0cm}{!}{\includegraphics{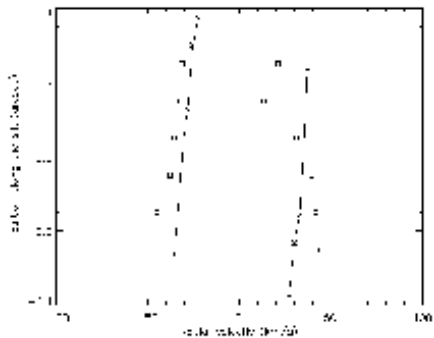}}
\caption{Mz~3 STIS central region data fitted to gaussians ($\Box$) and lobes model
($\times$).} 
\label{F15}
\end{figure*}

Given the complex apparent geometry of the lobes, we have decided to use an empirical numerical (not
analytical) description of their shape. This was done by marking the contour of the lobes on the HST
images, stretching it according to the inclination $i$ ({\it i.e.} the angle between the line of
sight and the symmetry axis of the structure), applying rotational symmetry, and scaling with
$v_{max}$ (the maximum velocity of the flow, corresponding in this case to the polar direction of
the Southern lobe) and the age-distance parameter $tD^{-1}$. This latter parameter contains the
inseparable dependence of the nebular size on the distance $D$ and on the kinematical age $t$,
assuming that the velocity of each particle of the flow has been constant since the original mass
ejection, and thus that the structure has grown in a self-similar way. 

A fair overall fit to the images and spectra was obtained (Figs.~\ref{F3} and \ref{F4}). The
best-fitting parameters, along with an estimate of the range of acceptable values, are listed in
table \ref{T1}. 

In spite of the good overall fit, spectra show some significant deviations from the model nebula at
specific positions. Some of them are likely to be related to intrinsic deviations from the
rotational symmetry that is indeed clearly seen both in the image and the spectra. But other
differences also exist. The most notable one is the excess of velocity seen in the spectra of the
polar protrusions as compared to the Hubble-flow model (see Fig.~\ref{F4}).

\begin{table}[!h] 
\begin{center} 
\begin{tabular}{c c c} 
Parameter & Value & Range\\ 
\noalign{\smallskip} 
\hline\hline 
\noalign{\smallskip} 
$tD^{-1} \ (year\ kpc^{-1})$ & $670$ & $(550-710)$\\ 
$v_{max} \ (km \ s^{-1})$ & $130$ & $(125-160)$ \\ 
$i (^\circ)$ & $73$ & $(71-76)$ \\ 
\noalign{\smallskip} 
\hline 
\end{tabular} 
\end{center} 
\label{T1} 
\caption{Best-fitting parameters for the lobes of Mz~3} 
\end{table} 
 
  \begin{figure*}
\center
\resizebox{4.0cm}{!}{\includegraphics{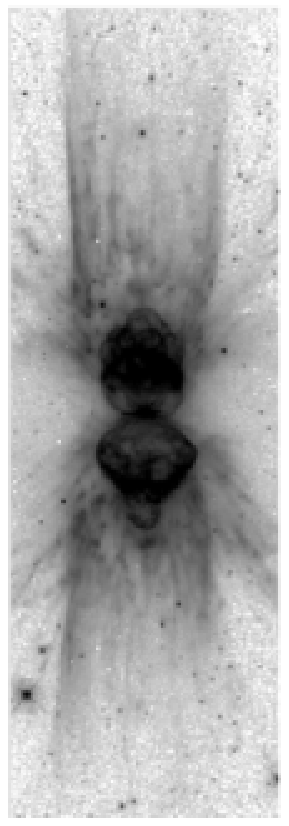}}
\resizebox{4.0cm}{!}{\includegraphics{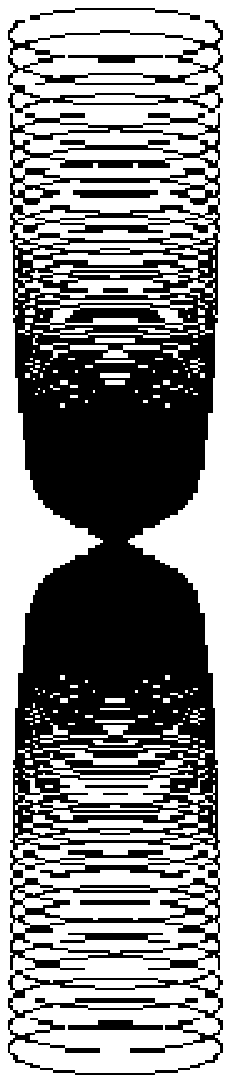}}
\caption{{\bf Left}: The H$\alpha$+[N{\sc ii}]\   ESO image of Mz~3. {\bf Right}: The best-fitting model for
the columns.} 
\label{F5}
\end{figure*} 
 
\begin{figure*}
\center
\resizebox{10.125cm}{!}{\includegraphics{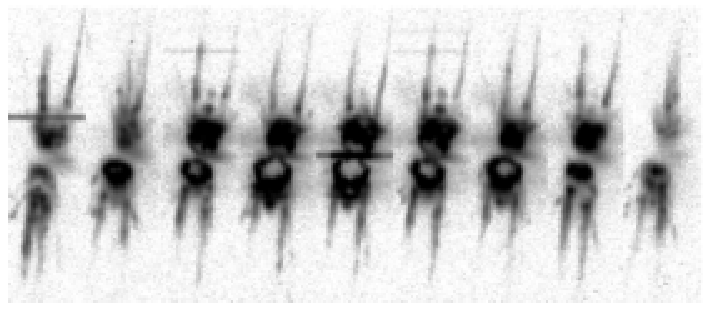}}
\resizebox{10.125cm}{!}{\includegraphics{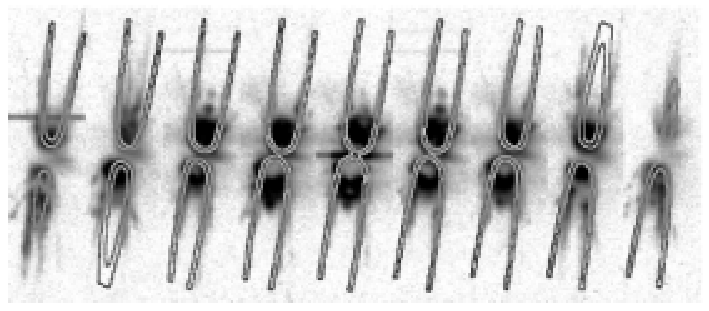}}
\caption{{\bf Top}: The CTIO [N{\sc ii}]\ echelle spectra. Frames
correspond to P.A.=$0^\circ$ and offsets from $-8''$ (E) to $+8''$ (W)
from left to right, in increments of 2''. {\bf Bottom}: The same
spectra with the spatio-kinematical model of the columns
superimposed. Model lines have been artificially broadened with
displaying purposes. {\bf Note}: Each frame is 120$''$ tall and 275 km
s$^{-1}$ wide. Northern side is up.}
\label{F6}
\end{figure*} 

The STIS long-axis spectrum allows us to probe deep into the circumnuclear regions. The [N{\sc ii}]\   line in
the STIS spectra of the innermost 0$''$.6 arcsec is resolved into two velocity components at all spatial
positions. A clear point-symmetrical pattern (the blueshifted component in the Northern direction
being much brighter than the redshifted one, and the opposite in the Southern side), is also clearly
present. The spectrum was binned to a 0$''$.1 resolution, and the two components fitted by Gaussians
at each spatial position. As shown in Fig~\ref{F15}, the position-velocity plot is consistent,
within the observational errors, with the adopted spatiokinematical model for the lobes. The
equatorial velocity of the model lobes, namely $\sim$10 km s$^{-1}$, appears therefore representative of the
expansion velocities of the [N{\sc ii}]\   emitting circumnuclear gas, which might be in a disk or torus-like
geometry. Both in H$\alpha$\  and [N{\sc ii}], the surface brightness in the nuclear region is
significantly enhanced with respect to the emission in the lobes. We
have checked whether this bright ionized core is spatially resolved in the
long axis
STIS spectra by measuring the Gaussian (spatial) centroids at the
wavelengths of the [N{\sc ii}]\ 6583~\AA line, and comparing them
with the center as defined by the nearby continuum emission (assuming that
that the continuum emission is produced very close to the central star).
This method allows to resolve the position of spectral features with an
accuracy much better than the instrumental resolution, as demonstrated by
Corradi et al. (\cite{Co99}) for the symbiotic nova HM Sge. In the case of
Mz~3, there is no evidence that the peaks of the [N{\sc ii}]\ emission are
displaced more than 0.01 arcsec from the position defined by the
continuum. This
puts an upper limit of 10$\cdot$$D_{kpc}$~a.u. for the size (along the
long axis of Mz~3) of the region where the enhanced [N{\sc ii}]\ nuclear emission
is produced. The geometry of the bright but unresolved circumstellar [N{\sc ii}]\ region  is not clear; however, the
kinematics of this
regions merge seamlessly  into the kinematics at the base of the lobes.  So they form a  large-scale coherent
kinematic feature despite their large difference  in surface brightness, leading us to speculate that the
circumstellar  gas may be a small dense disk or torus directly involved in the  collimation of the outflows.

\subsection{Columns}

In order to model the other outflows of Mz~3, we have used analytical description of their shapes. In the
following, we use the cylindrical coordinates $(z,\rho)$, with $z$ being the symmetry axis. Columns are
described using an
analytical formula that makes use of several parameters: a shape parameter, $n_{max}$, controlling
the curvature of the model especially near the central star; $a$, the initial relative width
($\frac{\rho_{max}}{z_{max}}$) of the model; $v_{min}$, the expansion velocity at the equator
($z=0$); as well as the previously defined $v_{max}$, $i$, and $tD^{-1}$. 
 
In formulae, the shape of the columns is described by 
 
$$
z= k_1 \ n^{(n^4+1)} 
$$
$$
\rho= k_2 \ n + r_{min}\, ,
$$ 

where $n$ goes from $0$ to $n_{max}$, and $r_{min}$ is the minimum distance to the central star, in
arcsec (the radius of the columns equatorial waist). $k_1$ and $k_2$ are normalization constants
defined as 

$$
 k_1=\frac{z_{max}}{n^{(n_{max}^4+1)}_{max}}
$$
$$
k_2=\frac{\rho_{max}}{n_{max}}\, .
$$

The best-fitting model image and spectra are shown in Figs.~\ref{F5} and \ref{F6}, and its
parameters are listed in table \ref{T2}. Again, except for the CTIO spectra corresponding to the
slit positions at offset $6''$ both to East and West, the fit is quite good for all images and
spectra. Besides, it nicely reproduces the single velocity and inclination values from the empirical model by
Meaburn \& Walsh (\cite{MW85}).
 
\begin{table}[!h] 
\begin{center} 
\begin{tabular}{c c c} 
Parameter & Value & Range\\ 
\noalign{\smallskip} 
\hline\hline 
\noalign{\smallskip} 
$tD^{-1} \ (year\ kpc^{-1})$ & $875$ & $(830-1000)$\\ 
$v_{max} \ (km \ s^{-1})$ & $300$ & $(250-425)$\\ 
$v_{min} \ (km \ s^{-1})$ & $4$ & $(0-10)$\\ 
$n_{max}$ & $1.5$ & $(1.45-1.53)$\\ 
$a$ & $0.19$ & $(0.15-0.19)$\\ 
$i (^\circ)$ & $76$ & $(76-78)$\\ 
\noalign{\smallskip} 
\hline 
\end{tabular} 
\end{center} 
\label{T2} \caption{Best-fitting parameters for the columns of Mz~3} 
\end{table} 
 
\subsection{Rays}

\begin{figure*}
\resizebox{8.0cm}{!}{\includegraphics{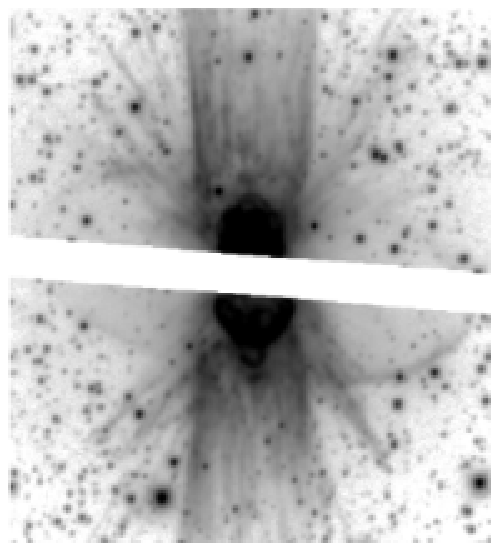}}
\resizebox{8.0cm}{!}{\includegraphics{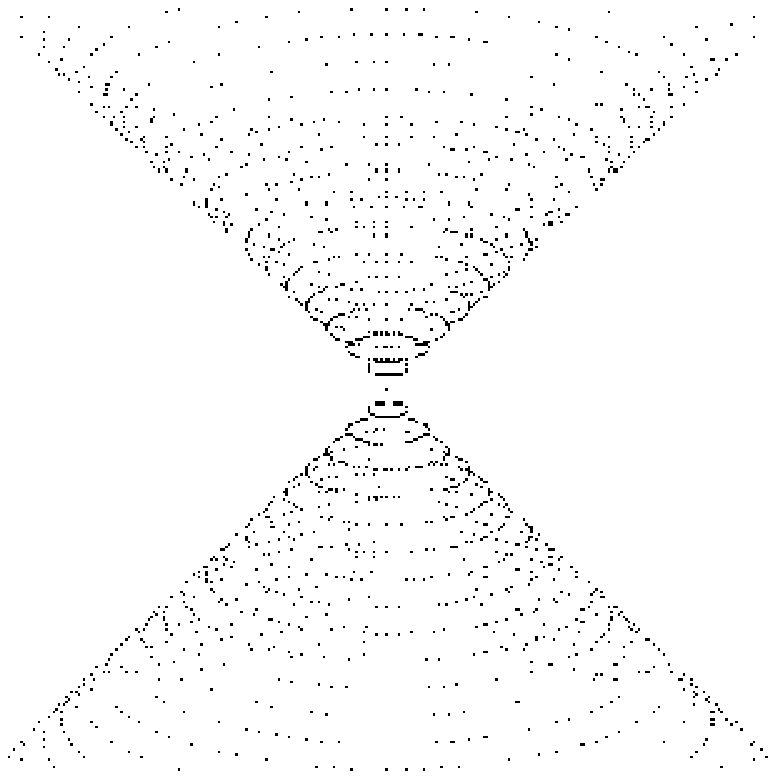}}
\caption{{\bf Left}: The H$\alpha$+[N{\sc ii}]\   ESO image of Mz~3. {\bf Right}: The best-fitting model for
the rays.} 
\label{F7}
\end{figure*} 
 
\begin{figure*}
\center
\resizebox{10.125cm}{!}{\includegraphics{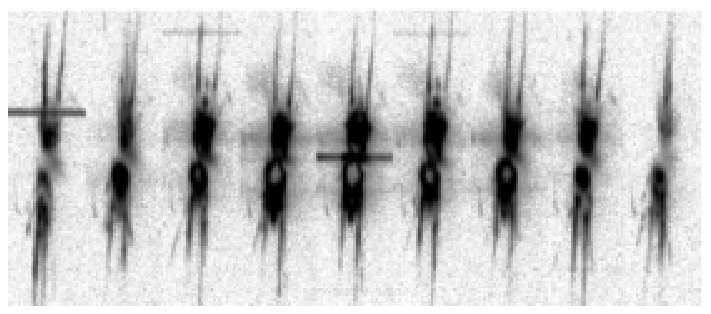}}
\resizebox{10.125cm}{!}{\includegraphics{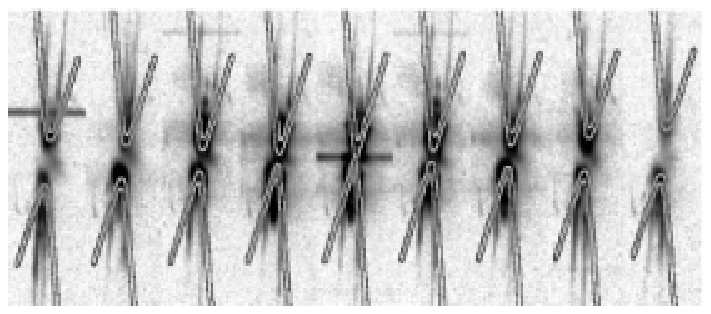}}
\caption{{\bf Top}: The CTIO [N{\sc ii}]\  echelle spectra. Frames correspond to P.A.=$0^\circ$
and offsets from $-8''$ (E) to $+8''$ (W) from left to right, in increments of 2$''$. {\bf Bottom}: The model of the rays
superimposed to
the CTIO spectra. As in previous cases, model lines have been broadened to improve the S/N ratio and get better displaying
results. {\bf Note}:
Each frame is 100$''$ tall and 480 km s$^{-1}$ wide. Northern side is up.} 
\label{F8}
\end{figure*} 

Rays were modeled as Hubble-like flows along the walls of a pair of reflected cones converging at the
central
star. The input parameters were $tD^{-1}$, $v_{max}$, $i$, and the complement $\theta$ to the cone
opening angle (or, in other words, the latitude of the walls of the cones from the equatorial plane). 

Again, a good analytical description of the geometry and outflow kinematics of the cones is shown in
Figs.~\ref{F7} and \ref{F8}, and the parameters of the model are given in table \ref{T3}. Note also that in
the aggregate the northern rays look somewhat different from the southern ones. For this reason they
were modeled separately. The results are similar on opposite sides of the star, but more uncertain for the
northern
side, where due to the lack of data, no paramter ranges are provided. The largest discrepancy are the expansion
ages, 1000 and 1600 years kpc$^{-1}$. We doubt that the
differences are significant or physically justifiable. Again, Meaburn \& Walsh's (\cite{MW85}) velocity and
inclination
empirical values are reliably reproduced by this fit.

\begin{table}[!h] 
\begin{center} 
\begin{tabular}{c c c} 
Parameter & Value & Range\\ 
\noalign{\smallskip} 
\hline\hline 
Southern Cone \\ 
\hline 
\noalign{\smallskip} 
$tD^{-1} \ (year\ kpc^{-1})$ & $1600$ & $(1400-1800)$\\ 
$v_{max} \ (km \ s^{-1})$ & $220$ & $(190-240)$\\ 
$\theta \ (^\circ)$ & $47$ & $(46-48)$\\ 
$i \ (^\circ)$ & $71$ & $(68-74)$\\ 
\noalign{\smallskip} 
\hline\hline 
Northern Cone \\ 
\hline 
\noalign{\smallskip} 
$tD^{-1} \ (year\ kpc^{-1})$ & $1000$: & \\ 
$v_{max} \ (km \ s^{-1})$ & $280$: & \\ 
$\theta \ (^\circ)$ & $47$:  & \\ 
$i \ (^\circ)$ & $74$:  & \\ 
\noalign{\smallskip} 
\hline 
\end{tabular} 
\end{center} 
\label{T3} \caption{Best-fitting parameters for the rays of Mz~3. ``:'' means uncertain.} 
\end{table}

\subsection{Chakram}

One of the oddest and unusual morphological features of Mz~3 is the large, faint, limb brightened ellipse that
seems centered on the nucleus. The plane of the ellipse is near, but clearly offset from the reflection
symmetry plane of the other features of Mz~3. The kinematics of this structure are unique among planetary
nebulae. They fail to show any velocity increase with radial offset from the nucleus, as do all of the other features
of Mz~3. Thus this is not presently a simple equatorial flow, though its motion seems to be strictly
radial (there is no sign of rotation that might suggest that the feature is dynamically stable). The kinematic
properties of this ellipse are very ordered and symmetric with respect to the nucleus, just like other
features of Mz~3. Hence the ellipse must have some sort of a historical link to the evolution of the star.

 \begin{figure*}[!]
\resizebox{8.0cm}{!}{\includegraphics{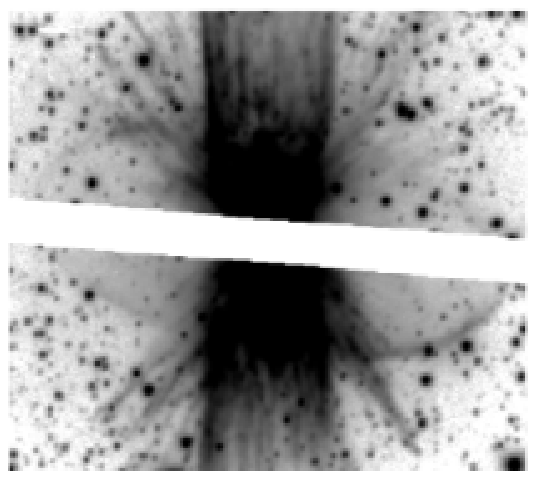}}
\resizebox{8.0cm}{!}{\includegraphics{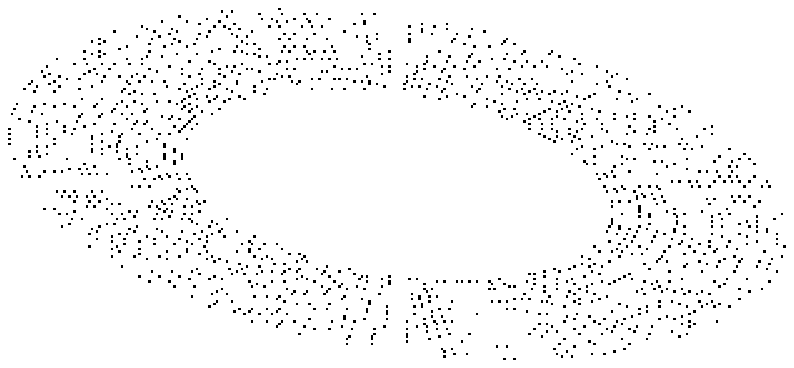}}
\caption{{\bf Left}: The H$\alpha$\ +[N{\sc ii}]\   ESO image of Mz~3. {\bf Right}: the adopted model for the
chakram.} 
\label{F9}
\end{figure*}

\begin{figure*}[!]
\center
\resizebox{9.0cm}{!}{\includegraphics{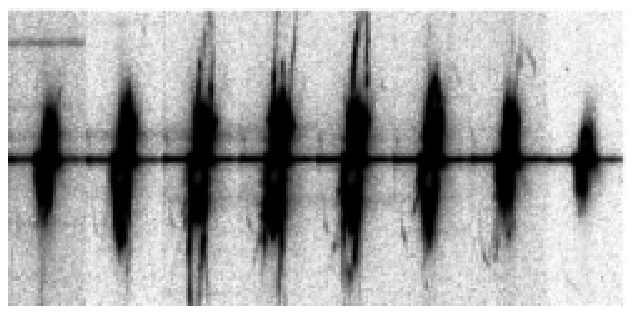}}
\resizebox{9.0cm}{!}{\includegraphics{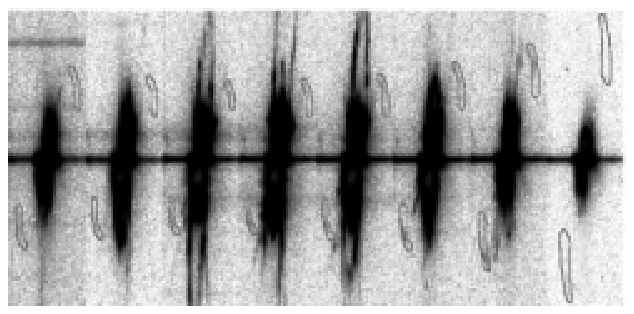}}
\caption{{\bf Top}: The ESO [N{\sc ii}]\  echelle spectra. Frames correspond, from left to right, to
P.A. from $-40^\circ$ to $+65^\circ$ in increments of $15^\circ$ and no offset. {\bf Bottom}: The same spectra with the
model of the chakram overlaid. Model lines have been exaggerately broadened so the reader can see both the model and the
spectra. {\bf Note}: Each
frame is 80$''$ tall and 480 km s$^{-1}$ wide. Northern side is up.} 
\label{F10}
\end{figure*}

\begin{figure*}[!]
\center
\resizebox{6.0cm}{!}{\includegraphics{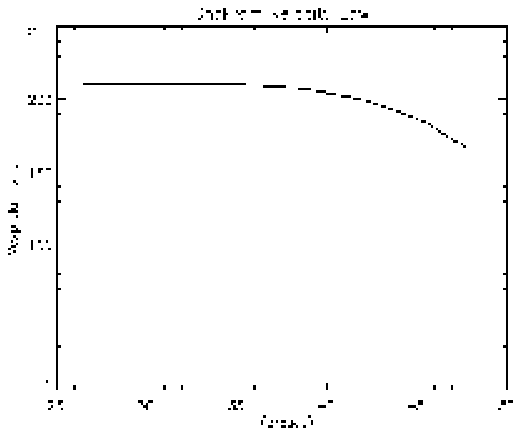}}
\caption{The velocity law of gas in the chackram, extracted and interpolated from ESO
spectra.} 
\label{F11}
\end{figure*} 

Finding a simple heuristic model for the ellipse was difficult. After considering the apparent morphology
and
the high-velocity component in the spectra, we decided
to model it as a broad and radially (but not ballistically) expanding annulus --a thin disk with an inner
hole-- centered on the star. We call this annulus the ``chakram''. The orientation of the symmetry axis of
this disk and its radial expansion profile are entirely free parameters of the model. 
A fit of the chakram to the images and spectra involves an empirical
determination of its inclination
{\it i} and its clockwise rotation angle $\tau$ with respect to the rest of structures. Given {\it i} and $\tau$, the
radial expansion
pattern was derived directly from the ESO spectrum at $P.A.=5^\circ$ using a
third-order polynomial.

In order to compute a kinematical age of this flow, a constant velocity from the central star to the
inner border of the chakram was assumed, by extrapolating the adopted expansion law (shown in
Fig.~\ref{F11}). From there on, it was calculated according to this law. 

Except for the external (East and West offsets) CTIO spectra, where it lacks of radial velocity (see
Figs.~\ref{F9} and \ref{F10}), and for the ESO spectrum at P.A.=$65^\circ$ where the chakram has
lower radial velocities than predicted by the model, the final fit to the chakram is also fair,
especially if we consider the very simple assumptions made. Note that spectra along position angles
near to the major apparent axis of the chakram are presently missing; they would be essential to
prove that the simple geometrical and kinematical description that we have chosen is indeed a good
description of the 
chakram. 

If confirmed by further spectra, a surprising result of our modeling would be that the disk-like
structure of the chakram is tangential (to within few degrees) to the walls of the cone describing
the rays (in other words, one diameter of the chakram would coincide with a ray). 
 
\begin{table}[!h] 
\begin{center} 
\begin{tabular}{c c c} 
Parameter & Value & Range\\ 
\noalign{\smallskip} 
\hline\hline 
\noalign{\smallskip} 
$tD^{-1} \ (year\ kpc^{-1})$ & $1100$ & -\\ 
$\tau (^\circ)$ & $9$ & $(8 - 10)$\\ 
$i \ (^\circ)$ & $-65$ & $-(63 - 68)$\\ 
\hline 
$r_{min} \ ('')$ & $26.5$ & $(24 - 30)$ \\ 
$r_{max} \ ('')$ & $47.8$ & $(45 - 54)$ \\ 
$v_{max} \ (km \ s^{-1})$ & $211$ & $(206 - 215)$ \\ 
$v_{min} \ (km \ s^{-1})$ & $168$ & $(164 - 170)$ \\ 
\noalign{\smallskip} 
\hline 
\end{tabular} 
\end{center} 
\label{T4} \caption{Best-fitting parameters for the chakram of Mz~3. The negative inclination means
that the chakram is inclined in the opposite direction to the other outflows.} 
\end{table}

 \begin{figure*}
\center
\resizebox{6.0cm}{!}{\includegraphics{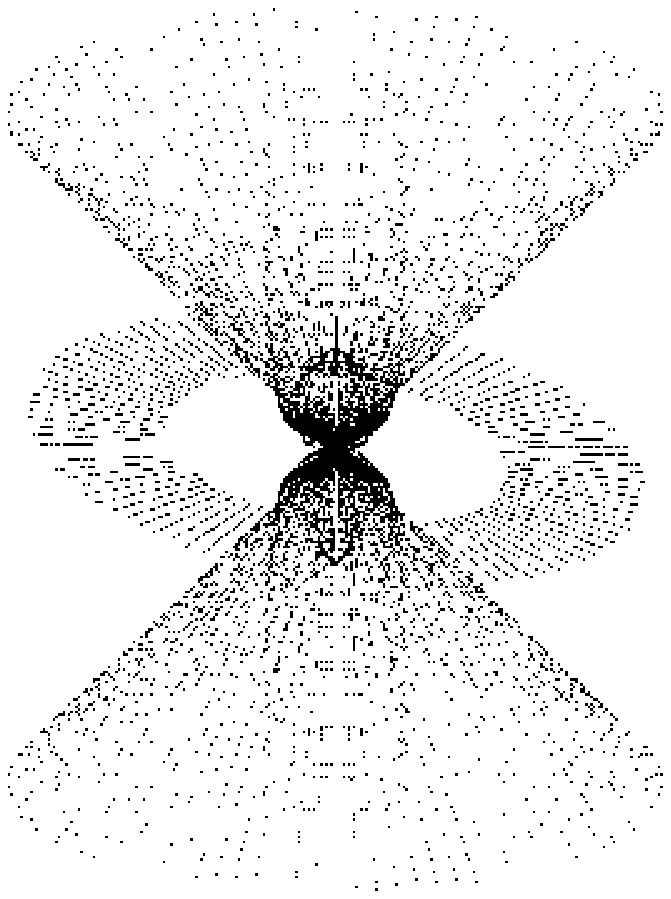}}
\resizebox{7.875cm}{!}{\includegraphics{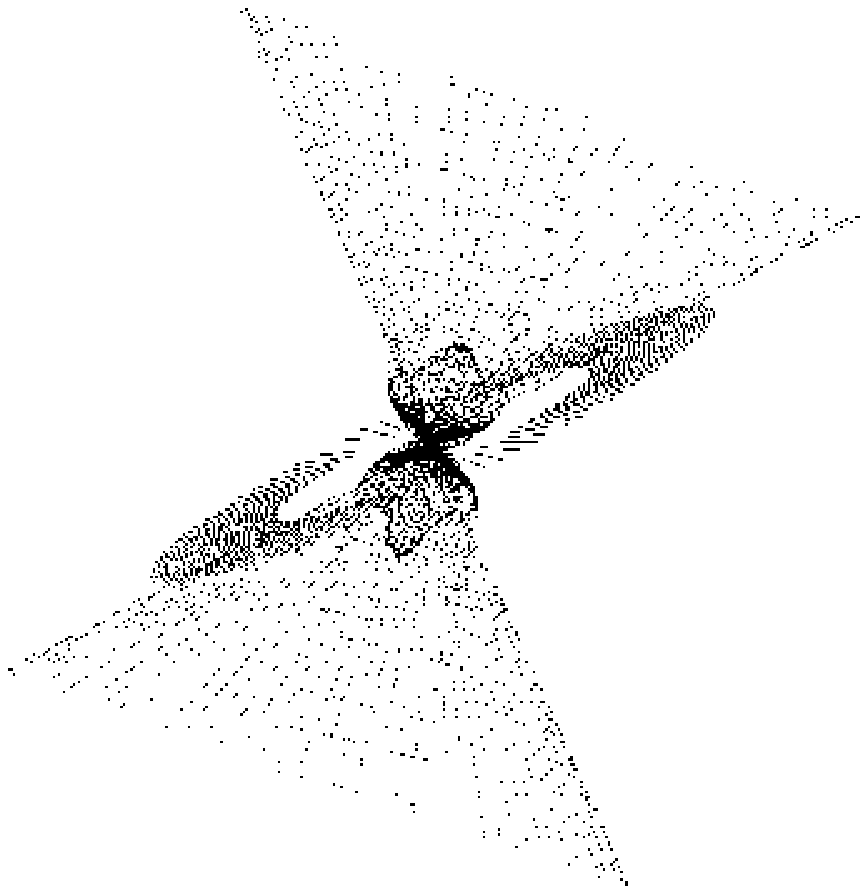}}
\caption{{\bf Left}: Mz~3 model projected on the plane of the sky. {\bf Right}: Transversal view of
Mz~3 model, i.e. after a $90^\circ$ rotation around $z$ axis.} 
\label{F14} 
\end{figure*}

\section{Discussion} 

Our analysis has revealed the existence of four distinct, high
velocity outflows, that we have named as lobes, columns, rays and
chakram. The overall geometry of the nebula is illustrated by the
sketch in Fig.~\ref{F14}. The first three features are adequately
described by co-axial flows following a Hubble-like expansion law. In
other words, each feature is the result of some sort of brief
formation process that is better characterized as eruptive and
structured rather than explosive and chaotic. Ignoring the chakram for
now, we find that Mz~3 fits into the same overall picture --brief
formation lasting perhaps a few hundred years followed by ballistic
growth-- found for other nebulae with symbiotic nuclei, M2-9 and
He2-104 (see Corradi \cite{Co04} for a more extended discussion). One
difference between Mz~3 from other similar nebulae is that its lobes,
columns, and rays have somewhat different ages from one another, with
the largest features generally being the oldest. This is expected: any
other pattern of formation and expansion predicts that the growing
features interact and almost certainly lose their integrity in the
process (shocks would convert the kinetic energy of their ordered
motion to heat). Nonetheless X-ray emission is seen interior to the
lobes of Mz~3 --see below.

On the other hand, we have also found that there are some significant
deviations from the Hubble law at specific positions. The lobes
present protrusions in their polar regions that show excess velocity
as compared to the simple Hubble-like flow (see Fig.~\ref{F4}). These
regions are also those that present X-ray emission (Kastner et
al. \cite{Ka03}), suggesting that strong hydrodymanical shaping might
be at work there. We have also found that the chakram flow does not
follow a Hubble law. According to the present data, this unique
structure is best described by a flat, broad expanding annulus
inclined so as that one of its diameters lies on the surface of the
pair of cones that contain the rays. The fit of the empirical model to
the data is compelling, albeit unique and very peculiar for planetary
nebulae.

According to the modeling, the gas forming the chakram would have been
continuously ejected during some 500 years kpc$^{-1}$, and as the
lower radial velocities are found in its most external regions where
surface brightness is also enhanced, it could have been decelerated at
its outer edge where the chakram accretes ambient material of lower
specific momentum (apparently without creating a leading
shock). Bujarrabal et al.  (\cite{Bu91}, \cite{Bu01}) do not detect
cold, dense molecular gas in Mz~3, so any gas upstream of the
expanding chakram must be hot and invisible, perhaps with a sound
speed larger than that of the speed of the leading edge of the
chakram. The absence of any Raleigh-Taylor instabilities or ruffles
along that leading edge seems to argue that the density of the
upstream gas is low. However, the apparent deceleration of the leading
edge requires that the mass of displaced ambient gas is comparable to
the mass of the chakram, perhaps the remnants of older winds ejected
by the star during the ascent of the AGB.

 The bizarre orientation of the chakram might be possibly
explained by the models of Garc\'\i a-Segura \& L\'opez (\cite{GL00}),
who find that misaligned structures (particularly in binary nuclei)
can be created as a point-symmetric feature in MHD models with the
magnetic collimation axis tilted with respect to the symmetry axis of
the bipolar wind outflow.

Given its complexity, it is not easy to conceive a scenario for the
formation of the nebula of Mz~3.  The presence of multiple outflows
with different collimation degrees requires a more articulated
mechanism rather than the standard interacting winds theory (Kwok et
al. \cite{Kw78}), where the interaction of the two winds that are
naturally produced in the past AGB evolution and in the post-AGB one
of single, solar-type stars, has been proved to be capable to form
bipolar outflows. It is likely that a more complex mass modulation
(possibly involving episodic fast winds) as well as other ingredients
like perhaps magnetic fields are needed to generate the various
outflows observed in Mz~3.

In this respect, the  morphology of the lobes, columns and rays
of Mz~3 resembles  very much model {\it U} in Garc\'\i a-Segura
et al.  (\cite{Ga99}), which would correspond to the mass loss from an
AGB star rotating close to its critical value, and with a strong
magnetic energy component. The expansion ages given there are also of
the same order of magnitude as determined in this work, and a linear
increase of velocity with distance from the central star, as generally
found in our work, is predicted for the jets produced by those
models. A deeper comparison with the present data is however prevented
by the lack, in the paper by Garc\'\i a-Segura et al. (\cite{Ga99}),
of a description of the kinematical properties of the model structures
that resemble the lobes, columns and rays of Mz~3.  It would be
extremely interesting to run new magnetohydrodynamical simulations
that are further constrained by the kinematical information recovered
in the present paper.

Useful information can be obtained by considering the ages of the
outflows. In spite of the errors intrinsic to the modeling adopted and
its simplistic assumptions, the present data would suggest a time
sequence in the ejection of the different outflows. The rays would be
the oldest ejecta, occurred some 1500 years ago (assuming a distance
of 1~kpc; in the literature one might find distances up to 2.5~kpc and
should scale ages consequently); then the columns, which would have
been ejected more than 500 years later; and finally the lobes, a few
hundred years younger. This would also suggest a sequence of
increasing collimation degree, which might be understood as due to an enhancement of
the magnetic field as the wind peels off the outer parts of the progenitor 
star, consistently with the results of Frank (\cite{Fr99}) and Franco et al. 
(\cite{FJ01}). The chakram is a highly peculiar
structure, and if the adopted modeling is confirmed to be correct by
additional data, would have been ejected during a lapse of time
overlapping with the lobes and columns.

Note, however, that the age differences are not big enough to rule out
that some of the flows (lobes and columns {\it or} columns and rays)
might have been produced approximately at the same time.

Finally, concerning the shaping of nested structures, consider the
lobes, for example. One might imagine two scenarios. The gas inside
the columns has a pressure which is either high or low when compared
with the ram pressure of the leading edges of the lobes expanding
inside them. In the former, the leading edges of the lobes should be
shocked, as it seems to happen, according to the limb brightening in
the images. However, there is no [O{\sc iii}], as it should, in the
hot recombination region behind the shock, just [N{\sc ii}]\ in the
warm recombination zone. In the latter case, the lobes would expand
against nearly vacuum except where the walls of the columns inertially
constrain them. Then their internal sound speed would be $\sim$100 km
s$^{-1}$, corresponding to a temperature of 10$^6$ K, which is
consistent with the X rays seen by Kastner et al. (\cite{Ka03}). The
expansion would cause the lobes to cool adiabatically, and their sound
speed would drop. Nonetheless the momentum of the leading edges would
result in ballistic motion (like the bullet of a rifle: the hot gases
that push the bullet cool long after the bullet moves ballistically
towards its target).

\section{Summary and conclusions} 

A spatiokinematical study of the bipolar nebula around Mz~3 has been
presented. Four different structures have been identified and named as
lobes, columns, rays and chakram. The first three have been modeled
following basic assumptions: axial symmetry and a Hubble-like,
self-similar radial ballistic growth. The chakram, a previously
unnoticed planar radially expanding outflow, appears to decelerate as
it expands, thus it has been modeled according to an empirical
velocity pattern.

The outflows have been found to expand at a maximum velocity that
range from $\sim$130 km s$^{-1}$ for the lobes and $\sim$300 km
s$^{-1}$ in the case of the columns. Similar but slightly different
dynamical ages (of the order of hundreds of years) with
increasing collimation degree have been found, and episodic eruptive
outflows proposed, with a brief discussion in the matter.

The chakram, on the other hand, has been found to have an odd
orientation, running along --and maybe braking against-- the walls of
the rays, with an expansion pattern that is not following a Hubble
law.

Recently we have been aware of an independent spatiokinematical study
of Mz~3 by Guerrero et al. (\cite{GMC}). They also use
spatiokinematical models for describing the different features of
Mz~3. Their derived kinematical parameters (ages, velocities and
inclinations) for the lobes, columns and rays appear to agree quite
well with our determinations.

\begin{acknowledgements}

We are grateful to Dr. Arsen Hajian for providing us with the CTIO
Echelle data; to Sean Doyle for the original, previous version of the
IDL code; and to Vincent Icke, for some general wisdom and criticism
which contributed to improve this paper.
 
\end{acknowledgements}

\end{document}